\documentclass{elsart}

\usepackage{graphicx}
\usepackage{dcolumn}

\begin{document}

\begin{frontmatter}

\title{Disorder effects at low temperatures in
La$_{0.7-x}$Y$_{x}$Ca$_{0.3}$MnO$_{3}$ manganites}

\author[poa]{M. A. Gusm\~ao\corauthref{mag}}\ead{magusmao@if.ufrgs.br},
\author[rio]{L. Ghivelder}, \author[rio]{R. S. Freitas},
\author[camp]{R. A. Ribeiro}, \author[camp]{O. F. de Lima},
\author[lon]{F. Damay}, \author[lon]{L. F. Cohen}

\address[poa]{Instituto de
F\'{\i}sica, Universidade Federal do Rio Grande do Sul, C.P.15051, Porto
Alegre, RS 91501-970, Brazil} 
\address[rio]{Instituto de F\'{\i}sica,
Universidade Federal do Rio de Janeiro, C.P.68528, Rio de Janeiro, RJ
21945-970, Brazil}
\address[camp]{Instituto de F\'{\i}sica ``Gleb Wataghin,'' UNICAMP, 
Campinas, SP 13083-970, Brazil}
\address[lon]{Blackett Laboratory, Imperial College, Prince Consort Road,
  London SW7 2AZ, United Kingdom}

\corauth[mag]{Corresponding author. Tel.: +55-51-33166481, FAX:
+55-51-33167286}

\begin{abstract}
With the aim of probing the effect of magnetic disorder in the
low-temperature excitations of manganites, specific-heat measurements were
performed in zero field, and in magnetic fields up to 9 T in polycrystalline
samples of La$_{0.7-x}$Y$_{x}$Ca$_{0.3}$MnO$_{3}$, with Y concentrations
$x=0$, 0.10, and 0.15. Yttrium doping yielded the appearance of a
cluster-glass state, giving rise to unusual low-temperature behavior of the
specific-heat. The main feature observed in the results is a strong
enhancement of the specific-heat linear term, which is interpreted as a
direct consequence of magnetic disorder. The analysis was further
corroborated by resistivity measurements in the same compounds.
\end{abstract}

\end{frontmatter}

The specific heat of manganites has attracted considerable interest in
recent years, yet several fundamental questions remain unanswered or only
loosely addressed. Starting with Ba- and Ca-doped LaMnO$_3$, initial
measurements at higher temperatures identified specific-heat anomalies
associated with phase transitions \cite{Ramirez,GhivJ3M}, while
low-temperature data \cite{GhivJ3M,Woodfield,Hamilton} showed a linear term
associated with an electronic contribution, and a cubic phonon
term. Subsequent studies were performed on a variety of samples, including
cation-deficient LaMnO$_{3 + \delta}$ \cite{Ghiv99}, Nd- \cite{Gordon} and
Pr- \cite{Lees,Hardy,smoly2,Raych} based compounds, or layered manganites
\cite{layer}. Recent studies \cite{Fisher,smoly3} addressed the issue of
isotope effect in the specific heat of manganites.

The existence of a spin-wave magnetic term in the specific heat of several
ferromagnetic (FM) manganite compounds is still a controversial issue. Spin
waves were not resolved in the specific heat of Ca-doped LaMnO$_{3}$
\cite{Hamilton,GhivJ3M} and Pr-based manganites \cite{Lees,smoly3}. They
were detected in neutron experiments on various compounds of the family, but
either coexisting with spin diffusion \cite{neutrons1} or presenting
excitation gaps that can be large enough \cite{neutrons2} to yield their
suppression at low temperatures.  A magnetic term was used to fit
specific-heat data for La$_{1-x}$Ca$_{x}$MnO$_3$ \cite{Okuda} and
La$_{1-x}$Sr$_{x}$MnO$_3$ \cite{Woodfield,Okuda}. In Ref.\ \cite{Woodfield}
it is clear that various sets of fitting parameters are possible, with very
different values of the spin-wave coefficient. Such ambiguities are not
uncommon, and high-resolution data combined with sophisticated fitting
techniques are often needed \cite{Fisher} in order to extract a FM spin-wave
contribution from specific-heat measurements. Apart from difficulties of the
fitting procedure, disorder might play a major role in suppressing these
excitations, as we will discuss below.

The most puzzling issue which emerged from previous investigations of the
specific heat is the observation of a large linear term in insulating
manganites \cite{Ghiv99,Gordon,Lees,smoly2}. The nature of the excitations
giving rise to this contribution remain unknown. Explanations based on
charge localization and/or spin disorder \cite{Ghiv99,smoly2} have been put
forward. An additional anomalous contribution, arising from excitations
following a dispersion relation $\Delta + \xi k^2$, has also been proposed
\cite{smoly2,smoly3}, and it has been suggested that these excitations are
related to charge ordering, although this interpretation still lacks
confirmation.

This paper presents low-temperature specific-heat and dc transport
measurements in La$_{0.7-x}$Y$_{x}$Ca$_{0.3}$MnO$_{3}$ manganites. Our main
goal is to address the origin of the large linear term of the specific heat
in samples where the carrier concentration is held constant but the carrier
mobility and magnetic properties are systematically modified by increasing
$x$. As Y is added in La$_{0.7}$Ca$_{0.3}$MnO$_{3}$, magnetic disorder
starts to play a major role in the system's properties. The compound evolves
from a nearly collinear ferromagnet at $x = 0$ to a magnetic cluster-glass
phase at $x = 0.15$ \cite{Freitas}, while a minimum that is observed in the
dc resistivity at low temperatures becomes more pronounced. Specific-heat
and resistivity data in this system, with and without an applied magnetic
field, provide invaluable insight into the effect of disorder in the
low-temperature excitations of manganite compounds.

The investigated compounds are polycrystalline samples of
La$_{0.7-x}$Y$_{x}$Ca$_{0.3}$MnO$_{3}$, with $x = 0$, 0.10, and 0.15,
prepared by solid-state reaction.  X-ray analysis confirmed a single-phase
orthorhombic perovskite structure. Scanning electron microscopy revealed an
average grain size of 3--6 $\mu$m, without significant variations from
sample to sample.  The specific-heat measurements were performed by
relaxation calorimetry, using a Quantum Design PPMS. The data were obtained
for temperatures between 2 and 20 K, and under applied magnetic fields up to
9 T.  The absolute accuracy of the system, checked against a copper sample,
is of the order of $\pm$2--3\%.  All samples had masses between 15 and 20
mg. The background signal, including the amount of Apiezon N grease used to
glue the sample on the platform, was recorded separately at all applied
fields, and subtracted from the measured heat capacity. The resistivity was
measured by a standard four-point technique. Magnetization and resistivity
results, measured in a wider temperature range (up to 300 K), can be found
in Ref.~\cite{Freitas}.

\begin{figure}[t]
\centering{\includegraphics{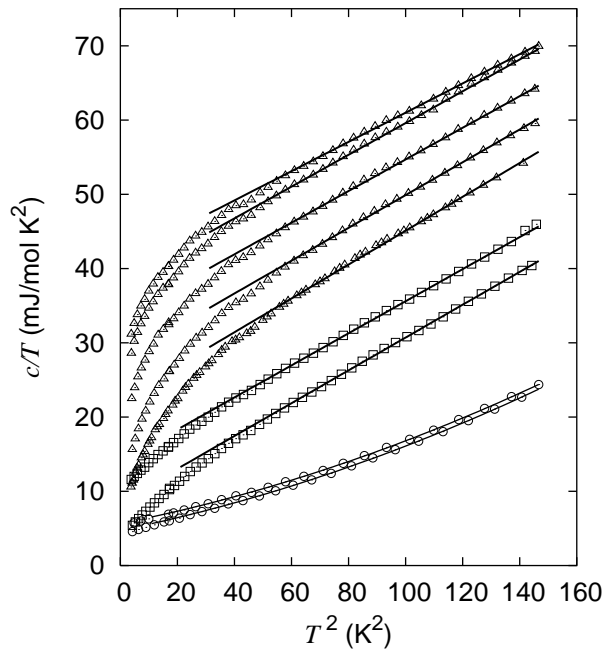}}
  \caption{Specific-heat measurements for $x=0.15$ (upper
    set -- triangles), 0.10 (middle set -- squares), and 0 (bottom set --
    circles). The highest and lowest curves of each set correspond to zero
    applied field and to $\mu_0 H = 9$ T, respectively. The intermediate
    curves for $x=0.15$ correspond to $\mu_0 H = 1$, 3, and 6 T,
    respectively from top to bottom. The solid lines are fittings to
    Eqs.~(\ref{eq:linear}) or (\ref{eq:phcorr}).}
  \label{fig:sph_all}
\end{figure}

The specific-heat data at low temperatures (2--12 K) for all the studied
samples and various applied magnetic fields are plotted as $c/T \times T^2$
in Fig.\ \ref{fig:sph_all}. We also show linear fittings to the data in the
upper temperature range.

The main features observed in those plots are: a large overall increase of
 the specific heat with yttrium content, a reduction of the specific heat
 with magnetic-field intensity, and a pronounced downturn of the data with
 decreasing temperature in the low-$T$ region for the yttrium-doped samples.
 The low-temperature downturn is much weaker in the yttrium-free
 sample, which also shows a slightly upward curvature of the data with
 increasing temperature.

Let us first concentrate on the ``high'' temperature range, where the
straight lines of Fig.~\ref{fig:sph_all} indicate a good fitting with the
standard expression
\begin{equation}
  \label{eq:linear}
  c(T) = \gamma T + \beta T^3 \;,
\end{equation}
where the $T^3$-term describes the phonon contribution, and the linear term
would be normally attributed to conduction electrons. However, this latter
interpretation has to be reviewed in the present case, as we will discuss
below. For the yttrium-free sample, we used
\begin{equation}
  \label{eq:phcorr}
  c(T) = \gamma T + \beta T^3 + \eta T^5 \;,
\end{equation}
including a $T^5$ correction to the phonon term, which accounts for the
observed upward curvature of the results. The best-fitting values of the
specific-heat coefficients $\gamma$, $\beta$, and $\eta$ are listed in
Table~\ref{tab:sph_coef}. The quality of the fitting is within 1\% in all
cases, and arbitrary variations the fitting temperature range above 7 K
yield change the obtained coefficients by less than 1\%. It is worth
mentioning that we have attempted to fit the specific heat data down to low
temperatures including a ferromagnetic $T^{3/2}$ term, as done in other
investigations \cite{Okuda}. However, in our case the quality of the fittings
are very poor below $\sim 4$ K, with an error above 20\% in some
cases. Furthermore, the magnitude of this contribution turns out to be
essentially insensitive to the applied magnetic field, in contrast to what
should be expected of magnetic excitations.

\begin{table}
\caption{\label{tab:sph_coef}Specific-heat coefficients of
  Eqs.~(\ref{eq:linear}) and (\ref{eq:phcorr}) for the fittings shown in
  Fig.~\ref{fig:sph_all}.}
\begin{center} \medskip
\begin{tabular}{cc|cr@{.}lcr@{.}lc} \hline
Y\% & $\mu_0H$ & \multicolumn{3}{c}{$\gamma$} & \multicolumn{3}{c}{$\beta$}
    & $\eta$ \\ & (T) & \multicolumn{3}{c}{(mJ/mol.K$^2$)} &
    \multicolumn{3}{c}{(mJ/mol.K$^4$)} & (mJ/mol.K$^6$) \\\hline
    \phantom{1}0 & 0 && 5&65 && 0&0766 & $3.49 \times 10^{-4}$ \\ & 9 &&
    4&73 && 0&0790 & $3.46 \times 10^{-4}$ \\ 10 & 0 && 14&0 && 0&216 & ---
    \\ & 9 && 8&61 && 0&221 & --- \\ 15 & 0 && 41&3 && 0&197 & --- \\ & 1 &&
    38&2 && 0&214 & --- \\ & 3 && 33&4 && 0&213 & --- \\ & 6 && 27&8 &&
    0&221 & --- \\ & 9 && 22&3 && 0&228 & --- \\ \hline
\end{tabular}
\end{center}
\end{table}

We now turn to a critical analysis of the results described above.  In
relation to the phonon terms, inspection of Table~\ref{tab:sph_coef} shows
that the absence of a $T^5$ correction in the yttrium-doped samples is
accompanied by an increase in the $T^3$ coefficient $\beta$, which indicates
a reduction of the Debye temperature. This is consistent with the generally
expected qualitative change of the phonon spectrum due to substitutional
impurities, i.\ e., localization of band-edge states, with the consequent
reduction of the effective band width. Suppression of the $T^5$ term may be
understood by the same argument, since the states that are detached from the
phonon band belong to the region where deviations from a linear dispersion
relation are more noticeable.

The absence of FM spin-wave contributions can be viewed as a consequence of
magnetic disorder in the samples. A cluster-glass structure has already been
observed in several manganites \cite{cglass,Ghiv99,Freitas} and related
compounds \cite{cobalt1,cobalt2}. Such a structure corresponds to
microscopic magnetic moments which align ferromagnetically within
finite-size clusters, with a random orientation of the cluster moments with
respect to each other. In this picture, long-wavelength spin waves would be
suppressed, and their effect not detected in low-temperature specific-heat
measurements. One could consider a spin-wave contribution with a
long-wavelength cutoff, or in other words, a gapped spin-wave spectrum with
a dispersion relation of the form $\omega(k) = \Delta + \xi
k^2$. Contributions from this kind of dispersion relation have been included
by other authors \cite{smoly2}, although attributed to excitations of a
different nature.  Despite the fact that part of the low-temperature
downturn of the specific heat observed in Fig.~\ref{fig:sph_all} can be
accounted for by such a term, its inclusion did not yield a reliable fitting
of the data. The same happens if one considers the existence of a gap in the
charge spectrum. As a general observation, these kind of terms tend to
compete with the linear one, so that weight can be shifted from one to the
other without affecting the quality of the fitting. Unfortunately, the
analysis of low-temperature specific-heat data is prone to ambiguities of
this sort, which have been noticed in other situations \cite{Hardy}. For
instance, a recent work by Hardy et al.\ \cite{Hardy2} reports a similar
downturn of the low-temperature specific heat for a system doped at the
manganese site. They try to fit the data including a phenomenological term
corresponding to some kind of gapped excitation, but various sets of fitting
parameters give equivalent results. In order to avoid these ambiguities, we
decided to concentrate in the most prominent features, which can be
unambiguously represented by the fittings to Eqs.~(\ref{eq:linear}) and
(\ref{eq:phcorr}) shown in Fig.~\ref{fig:sph_all}.

The large values of the linear coefficient $\gamma$ (see
Table~\ref{tab:sph_coef}), and the strong reduction of the specific heat in
the lower temperature range can also be interpreted as manifestations of
magnetic disorder. It is clear that the observed values of $\gamma$ are too
large to be attributed to conduction electrons. Furthermore, these values
are higher for lower-conductivity samples, which contradicts a naive
interpretation in terms of an enhanced density of states at the Fermi level.
The significant reduction of the $\gamma$ values under high applied magnetic
fields leads naturally to an interpretation in terms of excitations of
magnetic origin. Indeed, a linear behavior of the specific heat with
temperature is also observed in spin glasses \cite{sg}, which constitute a
classical example of magnetically disordered systems.  The temperatures and
coefficients are much higher in the present case, undoubtedly related to the
fact that here we are dealing with a system of large (cluster) magnetic
moments rather than a dilute spin system.  It is worth noticing that a large
linear term in the specific heat was previously observed in another
cluster-glass manganite compound \cite{Ghiv99}.

\begin{figure}[t]
\centering{\includegraphics[width=9cm,clip]{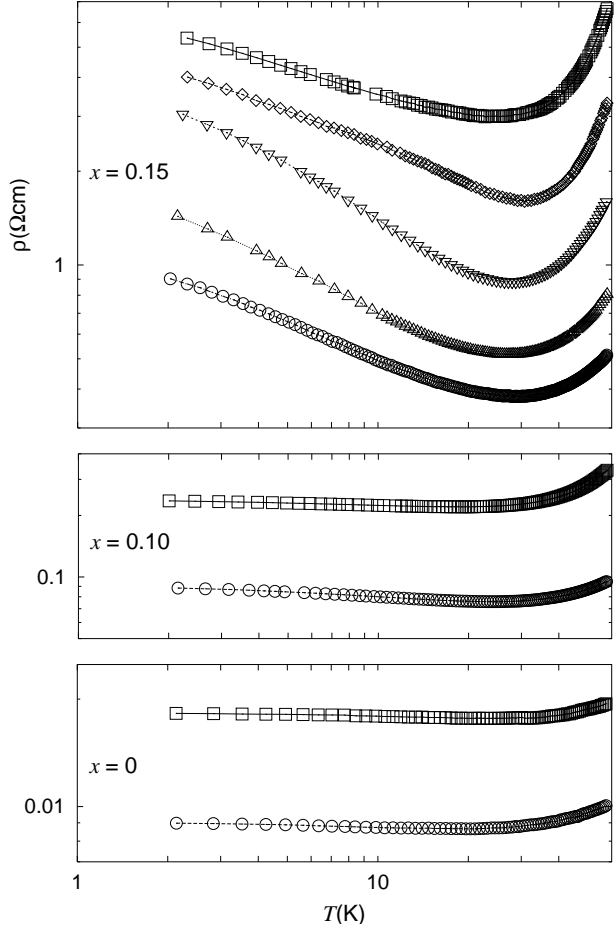}}
  \caption{Log-log plot of the low-temperature resistivity for all
    samples. The highest curve of each set (squares) is the zero-field
result.  The lowest curve of each set (circles) corresponds to $\mu_0 H = 9$
T. The intermediate curves for $x=0.15$ correspond to $\mu_0 H = 1$, 3, and
6 T, respectively from top to bottom. The lines joining the points are just
guides for the eye.}
  \label{fig:res}
\end{figure}

The low-temperature behavior of the electrical resistivity, shown in
Fig.~\ref{fig:res}, provides further clues. All samples show a shallow
minimum in the resistivity as a function of temperature, and the resistivity
values are significantly reduced by an applied magnetic field.  This is not
the usual (negative) colossal magnetoresistance shown by manganites, which
mainly occurs in a region around the Curie temperature. It is important to
notice that, although the resistivity grows as $T \to 0$, we {\em do not}
have carrier localization, since the conduction is not exponentially
activated. This can be seen in the log-log plots of Fig.~\ref{fig:res},
where the nearly straight lines at low temperature indicate a power-law
behavior that probably crosses over to a finite resistivity at $T=0$. The
large low-temperature magnetoresistance effect implies magnetic disorder
that is suppressed in a dc field. The increase of the absolute value of
resistivity with increasing $x$ supports this idea.

A similar low-temperature minimum of the resistivity was also observed in
manganite thin films \cite{minimum1}, and its origin has been discussed in
detail by Rozenberg et al.\ \cite{Rozenberg}. They argue against an
interpretation in terms of weak-localization effects, in which case the
field dependence would be much weaker, and interpret the resistivity
behavior as due to spin-polarized inter-grain tunneling conduction. We adapt
this model for our samples, and interpret the transport behavior in terms of
spin-polarized {\em inter-cluster} tunneling. With such a model, the
following expression for the low-temperature resistivity is proposed in
Ref.~\cite{Rozenberg}:
\begin{equation}
  \label{eq:rhograin}
  \rho(T,H) = \frac{\rho_U}{1+\varepsilon \langle \cos \theta_{ij} \rangle}
  \;,
\end{equation}
where $\theta_{ij}$ is the angle between the magnetizations of clusters $i$
and $j$, the brackets denote both thermal and configurational average,
$\rho_U$ is an intrinsic tunneling resistivity, independent of the
magnetization orientations, and $\varepsilon$ measures the degree of spin
polarization of the charge carriers in each cluster. The above expression
shows that the resistivity tends to be reduced if the inter-cluster
correlations are ferromagnetic, and greatly enhanced if they are dominantly
antiferromagnetic. The very definition of FM clusters implies that the AF
superexchange interactions will dominate over the double-exchange FM
coupling at the cluster boundaries, which is consistent with the resistivity
increase at low temperatures. Nevertheless, the system does not develop
long-range AF order, as the random distribution of clusters will tend to
frustrate the AF interactions.

The above discussion has shown that the same magnetically disordered cluster
structure accounts for both observed behaviors: enhanced linear term in the
specific heat, and low-temperature increase of the resistivity. This
consistency strengthens the reliability of this interpretation. One could
argue that the resistivity behavior could be due to inter-grain tunneling,
as the samples are polycrystalline. However, as we mentioned before, the
grain-size differences among the samples are very small, and cannot account
for the observed differences in the low-temperature resistivities.

In summary, we have studied the specific heat of
La$_{0.7-x}$Y$_{x}$Ca$_{0.3}$MnO$_{3}$ with $x=0$, 0.10, and 0.15, under
applied magnetic fields up to $\mu_0H=9$ T, in the temperature region
below 12 K.  The system is viewed as a collection of ferromagnetic clusters
that tend to align their moments antiferromagnetically, but their random
distribution prevents stabilization of long-range AF order, resulting in a
cluster-glass state. Low-energy excitations of this system yield a
contribution to the specific heat that is linear in $T$, with a coefficient
$\gamma$ that increases with $x$ and is strongly reduced by the presence of
an applied magnetic field, which suppresses magnetic disorder. At very low
temperatures, these magnetic excitations tend to be frozen out, reducing the
total specific heat, which shows a pronounced downturn in this temperature
region. Resistivity measurements, and the interpretation of those
measurements in terms of spin-polarized carrier tunneling between
antiferromagnetically correlated clusters, support the specific heat
observations.

To our knowledge, there are no theoretical predictions for the behavior of
the specific heat, or even further development of Eq.~(\ref{eq:rhograin})
for the resistivity, in this kind of model. Such predictions are needed for
a quantitative fit of the specific-heat results in the whole low-temperature
region, and to provide a better understanding about the nature of the
excitations occurring in this magnetically disordered system.

\section*{Acknowledgements}
 This work was supported by the Brazilian Ministry of Science and
Technology under Grant PRONEX/FINEP/CNPq No.\ 41.96.0907.00, by Conse\-lho
Nacional de Desenvolvimento Cient\'{\i}fico e Tecnol\'ogico (CNPq), and by
Funda\c{c}\~ao de Amparo \`a Pesquisa do Estado do Rio de Janeiro (FAPERJ).

\end{document}